\documentclass[12pt]{article}
\usepackage{amssymb,amsmath,amsbsy,amsthm,latexsym,graphicx}

\usepackage{authblk}

\theoremstyle{plain}

\theoremstyle{remark}

\theoremstyle{definition}

\newcommand{\Ref}[1]{(\ref{#1})}
\newcommand{\ii}{{\rm i}}

\renewcommand\Authfont{\scshape\normalsize}
\renewcommand\Affilfont{\itshape\small}
\author[1,*]{Jonas de Woul}
\author[2,\dag]{Jens Hoppe}
\author[2,\ddag]{Douglas Lundholm}
\author[1,\S]{Martin Sundin}

\affil[1]{Department of Theoretical Physics, Royal Institute of Technology (KTH)\newline
    106 91 Stockholm, Sweden \vspace{2mm}}
    
\affil[2]{Department of Mathematics, Royal Institute of Technology (KTH)\newline
    100 44 Stockholm, Sweden}

\title{\Large{\bf{A dynamical symmetry for supermembranes}}}
\date{}

\begin{document}

\maketitle

\let\oldthefootnote\thefootnote
\renewcommand{\thefootnote}{\fnsymbol{footnote}}
\footnotetext[1]{jodw02@kth.se}
\footnotetext[2]{hoppe@math.kth.se}
\footnotetext[3]{dogge@math.kth.se}
\footnotetext[4]{masundi@kth.se}
\let\thefootnote\oldthefootnote

\vspace{-1.2cm}

\begin{abstract}
A dynamical symmetry for supersymmetric extended objects is given.
\end{abstract}
Applying the Noether procedure to the supermembrane action \cite{BergshoeffSezginTownsend1987} in a light-cone formulation, and using work of Goldstone \cite{Goldstone}, who proved Lorentz invariance of the corresponding bosonic theory by finding a way to explicitly reconstruct $\zeta$ (usually denoted $x^-$) in terms of the transverse degrees of freedom, the generators of ``mixed'' rotations (involving the corresponding supersymmetrised version of $\zeta$) were given in \cite{deWitMarquardNicolai1990} (as expressions involving $\vec x$, $\vec p$, $\theta$, and, as in the bosonic theory, two discrete degrees of freedom: $\eta$ and $\zeta_0 = \int\rho\zeta d^2\varphi$) as
\begin{equation}
\label{Jiminus}
J_{i-} = \int\Bigl( x_i \mathcal{H}_S - \zeta_S p_i -\frac{\ii}{4\eta}\theta \gamma^{ik}\theta p_k- \frac{\ii}{8\eta}\epsilon^{rs}\partial_r x_j \partial_s x_k \theta \gamma^{ijk} \theta\Bigr) d^2\varphi,
\end{equation}
with
\begin{align}
\zeta_S &= \zeta_0 - \frac{1}{\eta} \int G^r (\varphi,\tilde \varphi)\Bigl(\frac{\vec p}{\rho}\cdot\tilde\partial_r \vec x + \frac{i}{2}\theta \tilde\partial_r \theta\Bigr)\rho(\tilde\varphi)d^2\tilde\varphi,\\
\mathcal{H}_S &= \frac{\vec{p}^2+\det(\partial_r \vec{x}\cdot\partial_s\vec{x})}{2\eta\rho} - \frac{\ii}{2\eta}\theta \gamma^i\epsilon^{rs}\partial_r x_i \partial_s \theta,
\end{align}
\begin{equation}
\int G^r(\varphi,\tilde\varphi)\rho d^2\varphi=0, \quad \tilde\nabla_r G^r(\varphi,\tilde\varphi)=\frac{\delta(\varphi,\tilde\varphi)}{\rho}-1,
\end{equation}
the $\vec{x}_i$, $\vec{p_j}$ canonically conjugate, and 
\begin{equation}
\{\theta_\alpha(\varphi),\theta_\beta(\tilde\varphi)\} = -\frac{\ii}{\rho}\delta_{\alpha\beta}\delta(\varphi,\tilde\varphi), \quad \{\eta,\zeta_0\} = 1,
\end{equation}
(all other dynamical Poisson brackets zero).

Recently, one of us \cite{Hoppe2010} found a dynamical symmetry for (bosonic) relativistic extended objects of any dimensionality. In the present note, it is pointed out that an analogous dynamical symmetry exists for supermembranes. The generators of internal transverse rotations,
\begin{equation}
\bar J_{ij} = \int \Bigl(x_i p_j -x_j p_i -\frac{\ii}{4}\theta \gamma^{ij}\theta \rho\Bigr)d^2\varphi - \Bigl(X_iP_j-X_jP_i\Bigr)+\frac{\ii}{4}\theta_0\gamma^{ij}\theta_0,
\end{equation}
\begin{equation}
X_i:=\int \rho x_i d^2\varphi, \quad P_i:=\int p_i d^2\varphi, \quad \theta_0 := \int \rho \theta d^2\varphi,
\end{equation}
and the purely internal (non zero mode) part of \Ref{Jiminus}, denoted by $\bar J_{i-}$, together with an additional (new) angular momentum formed out of dynamically generated Clifford variables, constitute a super-generalisation of the dynamical symmetry obtained in \cite{Hoppe2010}. Assuming that
\begin{equation}
\label{Full commutator of Jimnus}
\{ J_{i-} , J_{k-}\} = 0,
\end{equation}
the dynamical Poisson brackets $\{\bar J_{i-} ,\bar J_{k-}\}$ can be inferred by subtracting from \Ref{Jiminus} all terms involving zero modes:
\begin{align}
\bar J_{i-} &:= J_{i-} - J^{(0)}_{i-} - \tilde J_{i-},\\
J^{(0)}_{i-} &:= \Bigl(X_iH_S-\zeta_0 P_i\Bigr) - \frac{\ii}{4\eta}\theta_0 \gamma^{ik}\theta_0P_k,\\
\tilde J_{i-} &:= \frac{1}{\eta}\bar J_{ik}P_k-\frac{\ii}{2\eta}\theta_0\gamma^iQ,\\
Q_\beta &:=\int\Bigl(p_i\gamma^i_{\beta\alpha}\theta_\alpha+\frac{1}{2}\epsilon^{rs}\partial_r x_i \partial_s x_j \gamma^{ij}_{\beta\alpha}\theta_\alpha\Bigr)d^2\varphi - P_i\gamma^i_{\beta\alpha}\theta_{0\alpha}.
\end{align}
On the physical phase space, constrained by
\begin{equation}
\epsilon^{rs}\Bigl(\partial_r \vec p \cdot \partial_s \vec x + \frac{\ii}{2}\partial_r \theta \partial_s \theta\Bigr) = 0,
\end{equation}
we find
\begin{equation}
\{J^{(0)}_{i-} + \tilde J_{i-},J^{(0)}_{j-} + \tilde J_{j-}\} = - \frac{\mathbb{M}^2}{\eta^2} \bar{J}_{ij} - \frac{\ii}{4\eta^2}Q_\alpha\gamma^{ij}_{\alpha\beta}Q_\beta,
\end{equation}
with $\mathbb{M}^2$ the relativistically-invariant internal (squared) mass
\begin{equation}
\mathbb{M}^2 = 2\eta\int \mathcal{H_S} d^2 \varphi - \vec{P}^2.
\end{equation}
Assuming that 
\begin{equation}
\label{Commutator with Q}
\{\bar J_{i-},Q_\beta\}=0
\end{equation}
(which we deduced from a general argument, but better be checked explicitly as well), the cross terms vanish
\begin{equation}
\{\bar J_{i-},J^{(0)}_{j-} + \tilde J_{j-}\} + \{J^{(0)}_{i-} + \tilde J_{i-},\bar J_{j-}\} = 0.
\end{equation}
Hence we deduce that
\begin{equation}
\label{Jiminus commutator}
\{\eta\bar J_{i-},\eta\bar J_{k-}\} = \mathbb{M}^2\bar{J}_{ik} -Q_{ik} ,
\end{equation}
with
\begin{equation}
Q_{ik}:= - \frac{\ii}{4}Q_\alpha\gamma^{ik}_{\alpha\beta}Q_\beta.
\end{equation}
For $\mathbb{M}^2\neq 0$, one could also write this as
\begin{equation}
\label{Alternative Jiminus commutator}
\{\mathbb{M}_i,\mathbb{M}_k\} = \bar{J}_{ik} -\Sigma_{ik},
\end{equation}
where
\begin{equation}
\mathbb{M}_i:=\frac{\eta\bar J_{i-}}{\sqrt{\mathbb{M}^2}},
\end{equation}
and
\begin{equation}
\Sigma_{ik}:=- \frac{\ii}{4}\frac{Q_\alpha}{\sqrt{\mathbb{M}^2}}\gamma^{ik}_{\alpha\beta}\frac{Q_\beta}{\sqrt{\mathbb{M}^2}},
\end{equation}
is formed out of dynamically generated Clifford variables
\begin{equation}
\Sigma_{\beta}:=\frac{Q_\beta}{\sqrt{\mathbb{M}^2}}, \quad \{\Sigma_{\alpha},\Sigma_{\beta}\}=\delta_{\alpha\beta}.
\end{equation}
Observing that
\begin{align}
\{Q_{ij},Q_{kl}\} &= \Bigl(-\delta_{jk}Q_{il} +\delta_{ik}Q_{jl} - \delta_{il}Q_{jk} + \delta_{jl}Q_{ik}\Bigr)\mathbb{M}^2,\\
\{\bar J_{ik},\eta\bar J_{l-}\} &=-\delta_{kl}\eta\bar J_{i-}+\delta_{il}\eta\bar J_{k-},\\
\{Q_{ik},\eta\bar J_{l-}\}&=0,\\
\{\bar J_{ij}, Q_{kl}\} &=-\delta_{jk}Q_{il} +\delta_{ik}Q_{jl} - \delta_{il}Q_{jk} + \delta_{jl}Q_{ik},
\end{align}
one finally notes that, assuming \Ref{Full commutator of Jimnus} and \Ref{Commutator with Q}, $\eta \bar J_{i-}$ and $M_{ik}:=\mathbb{M}^2\bar J_{ik} - Q_{ik}$ (commuting with the supercharges $Q_\alpha$) generate the following  dynamical symmetry algebra for supermembranes
\begin{align}
\{M_{ij},M_{kl}\} &= \Bigl(-\delta_{jk}M_{il} +\delta_{ik}M_{jl} - \delta_{il}M_{jk} + \delta_{jl}M_{ik}\Bigr)\mathbb{M}^2,\\
\{\eta\bar J_{i-},\eta\bar J_{k-}\} &= M_{ik},\\
\{M_{ik},\eta\bar J_{l-}\} &=\Bigl(-\delta_{kl}\eta\bar J_{i-}+\delta_{il}\eta\bar J_{k-}\Bigr)\mathbb{M}^2.
\end{align}

\noindent{\bf Acknowledgement} 

We would like to thank E. Langmann for his kind support.

\end{document}